\documentclass[10pt,twoside]{article}
\usepackage{amssymb}

\usepackage{graphicx}

\begin{document}

\title{Gravitational waves in the generalized Chaplygin gas model}
\author{J. C. Fabris\thanks{%
e-mail: \texttt{fabris@cce.ufes.br}}, S. V. B. Gon\c{c}alves\thanks{%
e-mail: \texttt{sergio@cce.ufes.br}} \ and M. S. Santos\thanks{%
e-mail: \texttt{marcellesantos@yahoo.com}} \\
\\
\mbox{\small Universidade Federal do Esp\'{\i}rito Santo, Departamento
de F\'{\i}sica}\\
\mbox{\small Av. Fernando Ferrari s/n - Campus de Goiabeiras, CEP
29060-900, Vit\'oria, Esp\'{\i}rito Santo, Brazil}}
\date{\today}
\maketitle

\begin{abstract}
The consequences of taking the generalized Chaplygin gas as the
dark energy constituent of the Universe on the gravitational waves
are studied and the spectrum obtained from this model, for the flat case, is analyzed. Besides
its importance for the study of the primordial Universe, the gravitational waves represent an
additional perspective (besides the CMB temperature and
polarization anisotropies) to evaluate the consistence of the
different dark energy models and establish better constraints to
their parameters. The analysis presented here takes this fact into
consideration to open one more perspective of verification of the
generalized Chapligin gas model applicability. Nine particular
cases are compared: one where no dark energy is present; two that
simulate the $\Lambda$-CDM model; two where the gas acts like the
traditional Chaplygin gas; and four where the dark energy is the
generalized Chaplygin gas. The different spectra permit to
distinguish the $\Lambda$-CDM and the Chaplygin gas scenarios.
\vspace{0.7cm}

\par
KEYWORDS: dark energy, dark matter, gravitational waves
\par
\vspace{0.7cm}
\par
PACS numbers: 98.80.Bp, 98.65.Dx

\end{abstract}

\section{Introduction}
\label{intro}
The most recent results from type Ia supernovae
observations \cite{riess,permultter,tonry} and the cosmic microwave background anisotropies
detection \cite{wmap} have led cosmologists and astrophysicists to the
conclusion that most of the matter of the Universe interacts in a
repulsive manner (as an example through a negative pressure). There are
many candidates to describe this exotic fluid. The most natural candidate is
a cosmological constant, but it presents a discrepancy of $120$ order of
magnitude between the theoretical predictions and
the observational data \cite{cc}. A self-interacting scalar field, known as quintessence, is another
proposal to explain dark energy \cite{quint1}. However, it asks for fine tuning of microphysical
parameters in order to have a suitable potential term \cite{quint2}.
Among many other possibilities, the Chaplygin gas model has receveid recently special attention
\cite{pasquier,berto,fabris}.
The Chaplygin gas model is based on a perfect fluid whose pressure, besides to be
negative, varies inversely with density. One of the interesting aspects of this
fluid is a connection with branes in the context of string theories \cite{hoppe,ogawa,jackiw}. Some phenomenological
generalizations of this fluid have been proposed, leading to the so-called {\it generalized
Chaplygin gas model} \cite{berto}.
\par
The aim of this paper is to investigate the particular signatures of the
generalized Chaplygin gas, mainly in comparison to the standard $\Lambda$-CDM, in what
concerns gravitational waves. Special attention will be given to the spectral distribution of
energy density.
\par
The gravitational waves are very important to Cosmology.
They must have special signatures in the polarization \cite{paul1} of the CMB anisotropies \cite{paul2}.
Moreover, since gravitational waves has decoupled from matter already in the
deep early Universe, they can be a window to the primordial phase.
Even if gravitational waves (and consequently the polarization of the CMB photons
\footnote{There are two polarizations modes, called $E$ and $B$. The mode $E$ has
already been identified, but only the detection of the mode $B$ will allow to identify
gravitational waves \cite{kogut,kovac}.}) have
not been detected directly until now, great efforts are been done in this sense
and there is a hope that the next generation of experiment in space (LISA, Planck, etc.),
may allow this detection in perhaps ten years.
The GW
spectrum frequency range of observational interest extends from $10^{-18}Hz$ to $10^{10}Hz$
and the energy density spectrum is constrained by the CMB, in
units of the critical density $\Omega_{GW}$, as \cite{uzan}
\begin{equation}
\label{constr}
\frac{d\Omega_{GW}}{d \ln \nu}\Big|_{10^{-18}} \leq 10^{-10} \quad .
\end{equation}
This constraint will be used later in the evaluation of the spectra. Many works
have been made in order to identify specific signatures of cosmological models
in the spectra of gravitational waves, for example, in the case of quintessence model
\cite{uzan} and string cosmology \cite{ungarelli,sanchez,gasperini}.
\par
The generalized Chaplygin gas is characterized by the
equation of state
\begin{equation}
\label{stateeq}
 p=-\frac{A}{\rho^\alpha}\quad ,
\end{equation}
where $A$ and $\alpha$ are constants such that $A \geq 0$ and $0
\leq \alpha \leq 1$. If the energy-momentum tensor conservation is
taken into account, the relation between the generalized Chaplygin
gas density and the scale factor $a$ becomes
\begin{equation}
\label{density}
 \rho = \left(A+\frac{B}{a^{3(\alpha + 1)}}
 \right)^{\frac{1}{\alpha+1}}\quad ,
\end{equation}
where $B$ is an arbitrary integration constant. The main properties of the
relation (\ref{density}) are: (i) it interpolates the cosmological constant
phase (when the scale factor is large) and a pressureless fluid
(for small values of $a$) phase; (ii) the sound velocity
associated with it goes from zero, for $A=0$, to the velocity of
light, being always positive.

We will analyze the gravitational waves spectrum obtained from the
generalized Chaplygin gas model, which is described in
section \ref{section_model}. In section \ref{section_h} we discuss general
properties of the gravitational waves differential equation in the
context of two important dark matter models: cosmological constant
and Chaplygin gas. The spectrum obtained from
numerical calculations on these cases are presented in
section \ref{section_spectrum} and analyzed in section \ref{conclusions} where
we also present our conclusions.

\section{Outline of the model}
\label{section_model}
We consider a flat, homogeneous and
isotropic Universe described by the Friedman-Robertson-Walker
metric, which may be written as
\begin{equation}
\label{metric}
 ds^2=c^2dt^2-a^2(t) \ g^{(0)}_{ij} \ dx^i dx^j \quad ,
 \ \ \ \ g^{(0)}_{ij} \equiv \delta_{ij}\quad ,
\end{equation}
and leads the Einstein's equations to assume the form
\begin{equation}
\label{einsteineq1}
 \left(\frac{\dot{a}}{a}\right)^2 =
 \frac{8\pi G}{3} \ (\rho_m + \rho_c)\quad ,
\end{equation}
\begin{equation}
\label{einsteineq2}
 \frac{\ddot{a}}{a}+2\left(\frac{\dot{a}}{a}\right)^2=-8\pi G
 \ (p_m + p_c)\quad ,
\end{equation}
where $a$ is the scale factor of the Universe, while $\rho_m$ and
$\rho_c$ are the pressureless fluid and the Chaplygin gas
densities. The pressures $p_m$ and $p_c$ of the fluids are related
to their densities by the equations of state $p_m=0$ and $p_c=-A /
\rho_c^\alpha$.

The option of working in a flat Universe is justified by
the recent data from the CMB measurements. The curvature of the Universe is
characterized by the parameter
$\Omega_k$ which is defined, in terms of the total observed density
$\rho_T$ and the critical density $\rho_{cr}$, as
$\Omega_k=1-\Omega_T$, $\Omega_T=\rho_T / \rho_{cr}$. The CMB data gives
$\Omega_k=0 \pm 0.06$ \cite{Lineweaver}.

If the fluids interacts only through geometry, the
energy-momentum tensor for each component is conserved and we get
\begin{equation}
\label{rhos}
 \rho_m=\frac{\rho_{m_0}}{a^3} \ , \
 \rho_c=\left(A+\frac{B}{a^{3(\alpha + 1)}}\right)^
 {\frac{1}{\alpha + 1}}\quad .
\end{equation}
We take the scale factor today as the unity, $a_0=1$, and thus
$\rho_{m_0}$ and $\rho_{c_0}=(A+B)^{\frac{1}{\alpha + 1}}$ are the
present densities of the fluids (The subscripts $_0$, according to
the current notation, indicate we are considering the present
values of these quantities). From this last equation we can express the
integration constant $B$ in terms of $A$,
$B=\rho_{c_0}^{\alpha+1}-A$. The Chaplygin gas density is then
rewritten as
\begin{equation}
\label{rho_ch}
 \rho_c=\rho_{c_0}
 \left[\bar{A} + \frac{(1 - \bar{A})}{a^{3(\alpha + 1)}}
 \right]^{\frac{1}{\alpha + 1}} \ , \
 \bar{A}=\frac{A}{\rho_{c_0}^{\alpha + 1}}\quad ,
\end{equation}
and the parameter $\bar{A}$ is connected to the sound velocity in
the gas, $v_s$, by the expression
\begin{equation}
 v_{s_0} = c \ \sqrt{\frac{\partial p_c}
 {\partial \rho_c} \Big|_{t_0}} = c \ \sqrt{\alpha \bar{A}}\quad ,
\end{equation}
\par
Taking the first of the equations (\ref{rhos}) ( which refers to
the pressureless fluid ) and the equation (\ref{rho_ch}), we use
the relation $\Omega_{i_0} = \rho_{i_0} / \rho_{cr}$ (where
$i=m,c$) in (\ref{einsteineq1}) to obtain
\begin{equation}
\label{adot}
 \frac{\dot{a}}{a} = H_0 \left[\frac{\Omega_{m_0}}{a^3} +
 \Omega_{c_0} \left(\bar{A} + \frac{1 - \bar{A}}{a^{3 (\alpha +1)}}
 \right)^{\frac{1}{\alpha+1}} \right]^{1/2}\quad ,
\end{equation}
\begin{equation}
\label{addot}
 \frac{\ddot{a}}{a} = H_0^2 \left[- \frac{1}{2} \frac{\Omega_{m_0}}{a^3}
 + \Omega_{c_0}  \left(\bar{A} + \frac{1 - \bar{A}}
 {a^{3(\alpha + 1)}} \right)^{\frac{1}{\alpha + 1}}  \left(1 -
 \frac{3}{2} \ \frac{1-\bar{A}}{a^{3(\alpha + 1)}} \left(\bar{A} +
 \frac{1 - \bar{A}}
 {a^{3(\alpha + 1)}} \right)\right) \right]\quad ,
\end{equation}
where the Hubble constant $H_0$ is defined by the expression
$H_0={\dot{a}}_0/a_0$. Since we are restricted to a flat
Universe, the fractions of pressureless matter and Chaplygin gas
today, $\Omega_{m_0}$ and $\Omega_{c_0}$, obey to the relation
$\Omega_{m_0} + \Omega_{c_0} = 1$.

With these two last equations we are able to write the GW
amplitude differential equation as a function of the observable
variables $H_0$, $a$, $\Omega_{m_0}$ and $\Omega_{c_0}$, and of
the Chaplygin gas parameters $\bar{A}$, $\alpha$. It's also
important to remark that if $\bar{A}=0$, the gas behaves like
the pressureless fluid (and the situation is the same as if we had
set $\Omega_{m_0}=1$) while, on the other hand, it behaves like
the cosmological constant fluid when $\bar{A}=1$ (and therefore we
can simulate the $\Lambda$-CDM scenario).

Among the many possible cases produced by the combinations of
parameters we chose a few important ones and classify them
according to the fractions of dark energy and matter and to the
kind of dark energy, as shown in table \ref{table_models}  below.

\begin{table}[!h]
\begin{center}
\begin{tabular}{|c||c|c|c|c|}
\hline\hline
&  &  &  &  \\[-7pt]
& $\bar{A}$ & $\alpha$ & $\Omega_{c_0}$ & $\Omega_{m_0}$
\\[2pt] \hline\hline
&  &  &  &  \\[-7pt]
Pressureless Fluid & -- & -- & 0 & 1 \\[2pt] \hline
&  &  &  &  \\[-7pt]
Cosmological Constant & 1 & -- & 0.96 & 0.04 \\[2pt] \cline{4-5}
&  &  &  &  \\[-7pt]
&  &  & 0.7 & 0.3 \\[2pt] \hline
&  &  &  &  \\[-7pt]
Generalized Chaplygin Gas & 0.5 & 1 & 0.96 & 0.04 \\[2pt] \cline{4-5}
&  &  &  &  \\[-7pt]
&  &  & 0.7 & 0.3 \\[2pt] \cline{3-5}
&  &  &  &  \\[-7pt]
&  & 0.5 & 0.96 & 0.04 \\[2pt] \cline{4-5}
&  &  &  &  \\[-7pt]
&  &  & 0.7 & 0.3 \\[2pt] \cline{3-5}
&  &  &  &  \\[-7pt]
&  & 0 & 0.96 & 0.04 \\[2pt] \cline{4-5}
&  &  &  &  \\[-7pt]
&  &  & 0.7 & 0.3 \\[2pt]
\hline\hline
\end{tabular}
\end{center}
\caption{{\footnotesize Models of interest for gravitational waves
power spectrum analysis.}}
\label{table_models}
\end{table}

In the first case no dark energy is present,
while the other eight cases reflect the situation where the dark
energy constitutes all the non-visible matter (that means, no dark
matter exists and $\Omega_{m_0}=0.04$) or the one where the dark
matter and the visible matter altogether represent 30\% of the
total energy density ($\Omega_{m_0}=0.3$).

After establishing the characteristics of the studied models, we
start in section \ref{section_h} the description of the behavior of gravitational
waves due to different cosmic fluids contents, namely the cosmological
constant and the generalized Chaplygin gas.

\section{GW equation in Chaplygin gas and $\Lambda$-CDM models}
\label{section_h}

Cosmological gravitational waves are obtained by means of a small
correction $h_{ij}$ on equation (\ref{metric}), which represents
the metric. Hence, the tensor $g^{(0)}_{ij}$, related to the
unperturbed metric, is replaced by $g_{ij} = g^{(0)}_{ij} +
h_{ij}$ and the resulting expression is \cite{weinberg,lifschitz}:
\begin{equation}
\label{gw1}
 \ddot{h} - \frac{\dot{a}}{a} \dot{h}
 + \left( \frac{k^2}{a^2} - 2 \frac{\ddot{a}}{a} \right) h = 0\quad ,
\end{equation}
where $k$ is the wave number times the velocity of light ($k=2 \pi
c/\lambda$), the dots indicate time derivatives and we have written, $h_{ij}(t,\vec x) =  h(t)Q_{ij}$, where $Q_{ij}$ are the eigenmodes of the Laplacian operator such
that $Q_{ii} = Q_{ki,k} = 0$.

Performing a variable transformation, from time to the scale
factor $a$, and representing the derivatives with respect to $a$
by primes, equation (\ref{gw1}) assumes the form
\begin{equation}
\label{gw2}
 h'' + \left( \frac{\ddot{a}}{\dot{a}^2} - \frac{1}{a} \right)h'
 + \frac{1}{\ddot{a}^2} \left( \frac{k^2}{a^2}
 - 2 \frac{\ddot{a}}{a} \right)h = 0 \quad .
\end{equation}
By using the background equations (\ref{adot})
and (\ref{addot}), or similar ones concerning to other fluids,
into (\ref{gw2}), one can easily express $h$ in terms of the
parameters of the Chaplygin gas or any other model.

\subsection{Chaplygin Gas}
\label{subsection_ch}
Let us perform the last operation mentioned
above and find $h$ as a function of the redshift $z$, with the following steps:
(i) use the fairly known relations $1+z=\frac{a_0}{a}$, $a_0=1$;
(ii) perform a second variable changing (from $a$ to $z$); and
(iii) take back the dots to indicate, from now on, the new
integration variable. These operations result in
\begin{equation}
\label{gw_ch}
 \ddot{h} + \left[\frac{2}{1+z}+\frac{3}{2}(1+z)^2 \frac{f_1}{f_2}
 \right]\dot{h} + \left[\frac{k^2}{f_2}-\frac{2}{(1+z)^2}+3(1+z)
 \frac{f_1}{f_2} \right]h = 0 \quad ,
\end{equation}
\begin{equation}
\label{f1}
 f_1=\Omega_{m_0}+\Omega_{c_0}(1-\bar{A})(1+z)^{3\alpha}
 \left[\bar{A}+(1-\bar{A})(1+z)^{3(\alpha+1)} \right]^
 {\frac{1}{\alpha+1}-1} \quad ,
\end{equation}
\begin{equation}
\label{f2}
 f_2=\Omega_{m_0}(1+z)^3 + \Omega_{c_0} \left[\bar{A}+
 (1-\bar{A})(1+z)^{3(\alpha+1)} \right]^{\frac{1}{\alpha+1}}\quad ,
\end{equation}
where $k$ has been redefined to absorb the Hubble constant, i.e., $k =
2\pi c H_0 / \lambda$. Therefore, $h$ is a dimensionless
quantity.

The solutions $h(z)$ of the differential equation oscillates with
greater amplitudes as $z \to 0$: these
amplitudes are increasing from the decoupling era up to the
present.

Equations (\ref{gw_ch}-\ref{f2}) above may be used to perform all
the cases of interest mentioned before. Setting $\Omega_{m_0}=1$
and $\Omega_{c_0}=0$, for example, we have the pressureless fluid
equation and for the cosmological constant we set $\bar{A}=1$.

\subsection{Cosmological constant}
\label{subsection_cc}

In order to establish the GW equation for the cosmological
constant fluid, we use the fact that the Chaplygin gas reproduces
its behavior if $\bar{A}=1$, for any value of $\alpha$, as stated before. Hence, the
equations (\ref{f1}) and (\ref{f2}) become
\begin{equation}
\label{f1_cc}
 f_1=\Omega_{m_0}\quad ,
\end{equation}
\begin{equation}
\label{f2_cc}
 f_2=\Omega_{m_0}(1+z)^3+\Omega_{\Lambda_0}\quad .
\end{equation}

The resulting equation is easier to integrate than the one
obtained before and presents the same qualitative behavior with respect to the
redshift. Another important feature to be mentioned is the
existence of quite simple analytical solutions, which permit us
write $h(z)$ in terms of a combination of sines and
cosines:
\begin{equation}
\label{solh_cc}
 h(z)= \frac{\sqrt{2/\pi}}{k^{3/2}(1+z)^3}
 \Big[(k(1+z)C_1+C_2)\cos{[k(1+z)]}+
 (k(1+z)C_2-C_1)\sin{[k(1+z)]} \Big]\quad .
\end{equation}

The arbitrary constants $C_1$ and $C_2$ remain undetermined since
the set of initial conditions is unspecified. Nevertheless, it is
important to remark that, for a fixed $k$, the argument $\theta$
of the cosine and sine functions in (\ref{solh_cc}) is
$\theta=k(1+z)$ and so the oscillation frequency decreases when
$z$ tends to 0.

\begin{figure}[!h]
\centering
\includegraphics[scale=0.4]{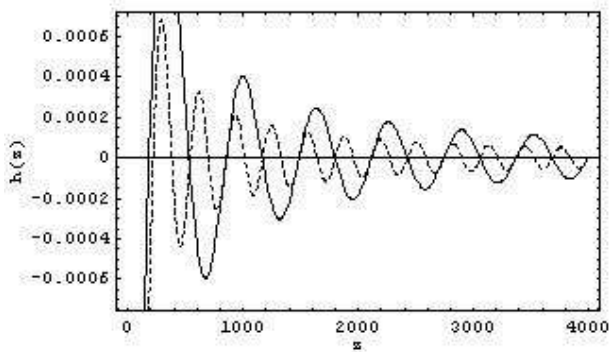}
\caption{{\footnotesize Gravitational waves curves for
$\Lambda$-CDM model. The dash line refers to $k=1/100$ and the
solid line to $k=1/50$.}} \label{fig_h_cc}
\end{figure}

Figure (\ref{fig_h_cc}) illustrates the case where
$\ddot{h}(z_i)=\dot{h}(z_i)=10^{-6} , z_i=4000$ are the initial
conditions for two different values of $k$.

The behavior described is evident and one can also observe that
the amplitude grows with time (i.e., as $z \to 0$), as we have
remarked for the Chapligyn gas. This result implies
that initial small primordial fluctuations on the gravitational
field are amplified.

\section{GW spectra}
\label{section_spectrum}

The power spectrum of gravitational
waves, defined as \cite{uzan}
\begin{equation}
\label{powerspectrum}
 \frac{d\Omega_{GW}}{d ln{\nu}}=|h_0(\nu)|\nu^{5/2}\quad ,
\end{equation}
(where $h_0(\nu)=h(0)$ and $\nu=H_0 k / 2\pi$), is generally
obtained directly from the solutions of (\ref{gw_ch}).
In the particular case of the $\Lambda$-CDM model, an analytical
result is also possible, as shown in (\ref{solh_cc}), and this fact
may be used to verify the accuracy of the calculation and the
applicability of the algorithm.

\begin{figure}[!h]
\centering
\begin{minipage}[t]{0.47\linewidth}
\includegraphics[width=\linewidth]{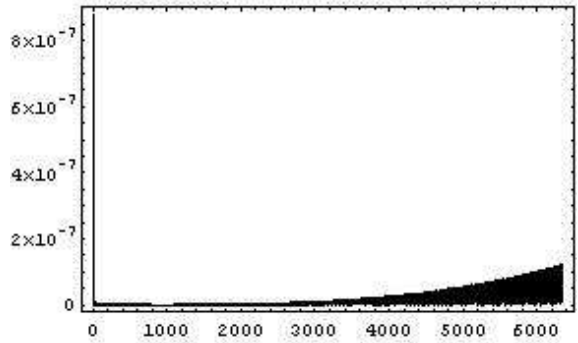}
\caption{{\footnotesize Graphic $(d\Omega_{GW}/dln\nu)$ X $(\nu\times
10^{-18}Hz)$
for the model where $\Omega_{c_0}=0$ and $\Omega_{m_0}=1$}}
\label{espectro_5}
\end{minipage} \hfill
\end{figure}

\begin{figure}
\begin{minipage}[t]{0.47\linewidth}
\includegraphics[width=\linewidth]{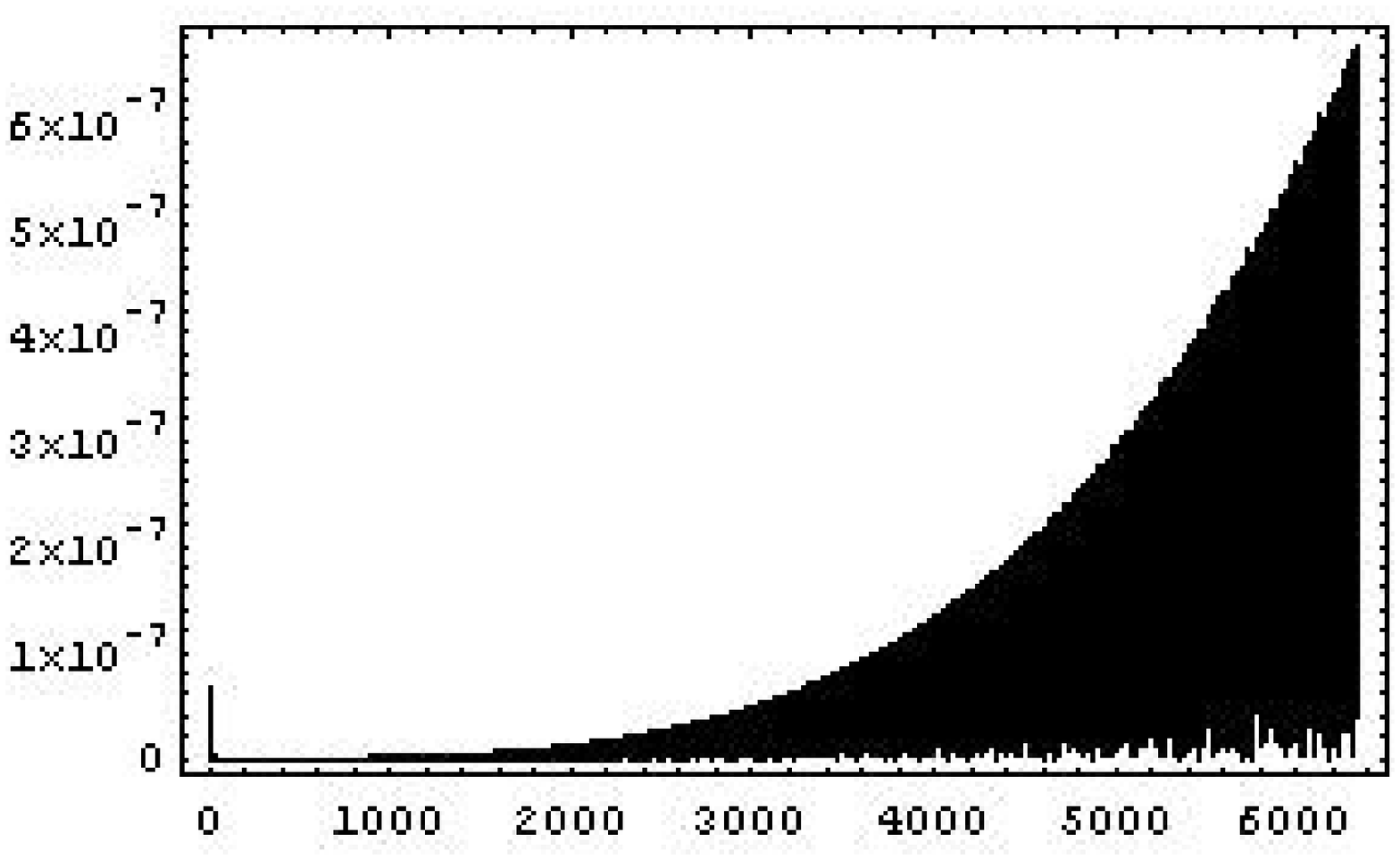}
\caption{{\footnotesize Graphic $(d\Omega_{GW}/dln\nu)$ X $(\nu\times
10^{-18}Hz)$
for the model where $\bar{A}=1$, $\Omega_{c_0}=0.96$ and
$\Omega_{m_0}=0.04$}} \label{espectro_14}
\end{minipage} \hfill
\begin{minipage}[t]{0.47\linewidth}
\includegraphics[width=\linewidth]{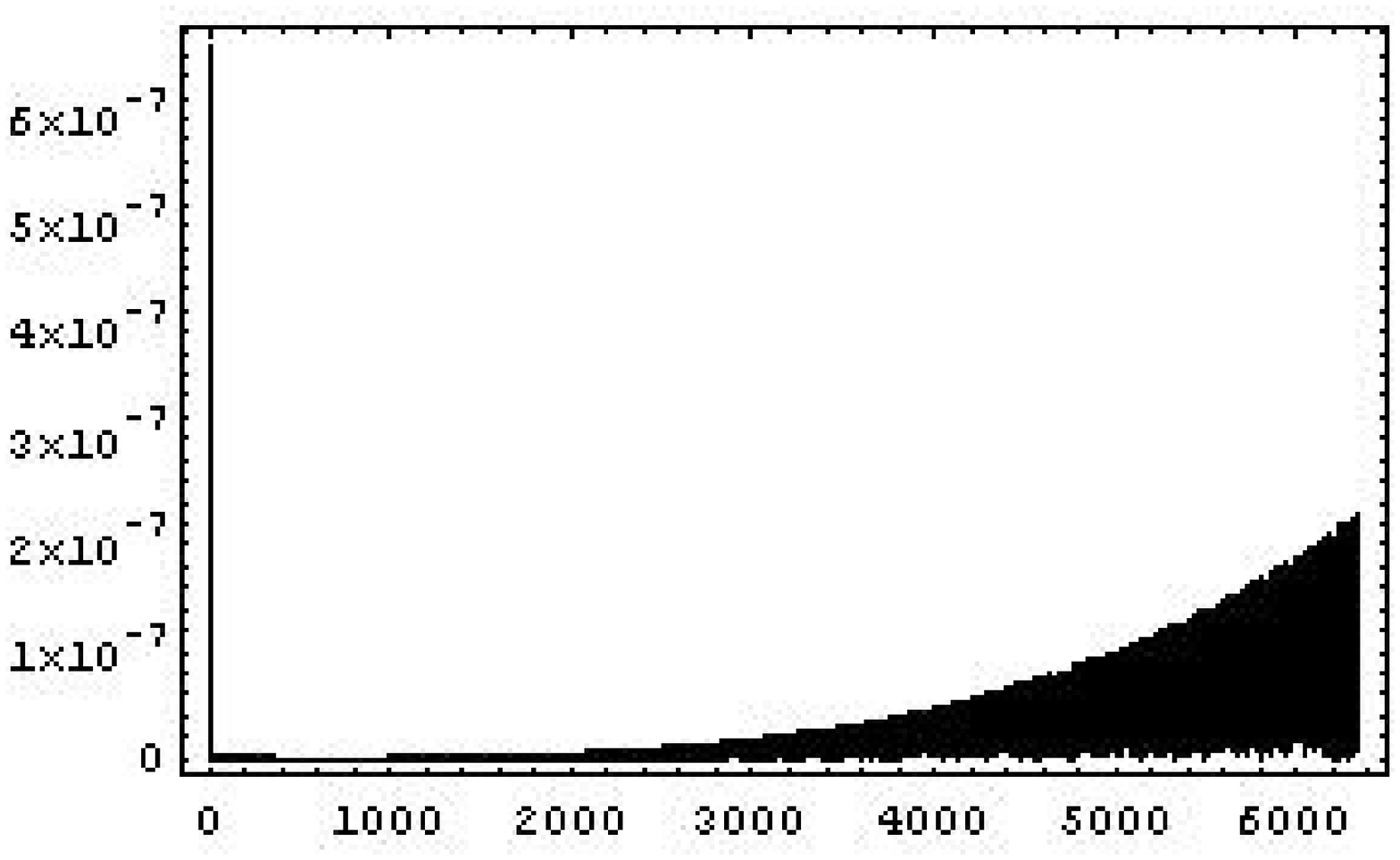}
\caption{{\footnotesize Graphic $(d\Omega_{GW}/dln\nu)$ X $(\nu\times
10^{-18}Hz)$
for the model where $\bar{A}=1$, $\Omega_{c_0}=0.7$ and
$\Omega_{m_0}=0.3$}} \label{espectro_15}
\end{minipage} \hfill
\end{figure}

\begin{figure}
\begin{minipage}[b]{0.47\linewidth}
\includegraphics[width=\linewidth]{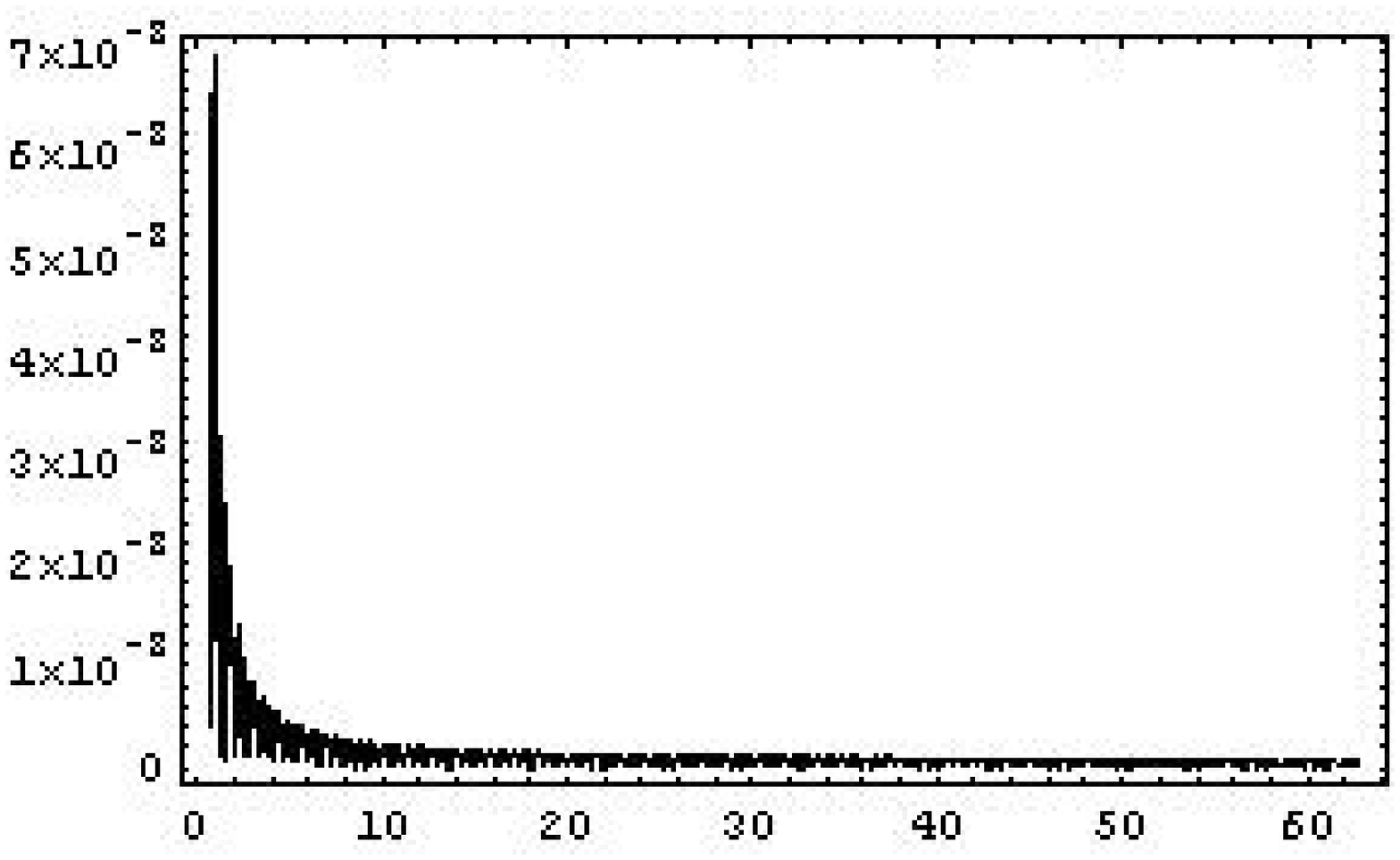}
\caption{{\footnotesize Graphic $(d\Omega_{GW}/dln\nu)$ X
$(\nu\times 10^{-18}Hz)$ for the model where $\bar{A}=1$,
$\Omega_{c_0}=0.96$ and $\Omega_{m_0}=0.04$ for small values of
$k$.}} \label{fig_5}
\end{minipage} \hfill
\begin{minipage}[b]{0.47\linewidth}
\includegraphics[width=\linewidth]{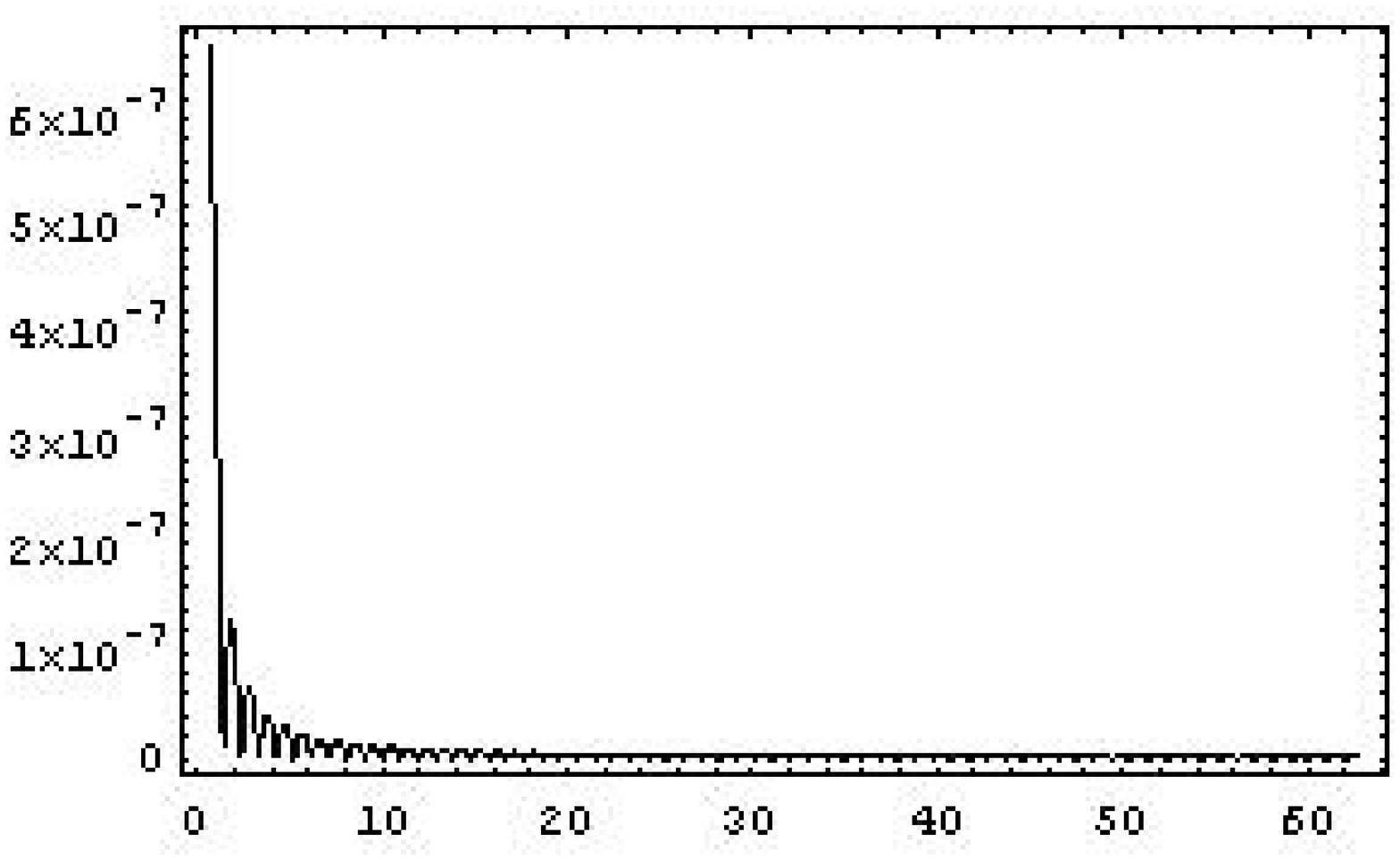}
\caption{{\footnotesize Graphic $(d\Omega_{GW}/dln\nu)$ X $(\nu\times
10^{-18}Hz)$
for the model where $\bar{A}=1$, $\Omega_{c_0}=0.7$ and
$\Omega_{m_0}=0.3$ for small values of $k$.}} \label{fig 1}
\end{minipage} \hfill
\end{figure}

\begin{figure}[!h]
\begin{minipage}[b]{0.47\linewidth}
\includegraphics[width=\linewidth]{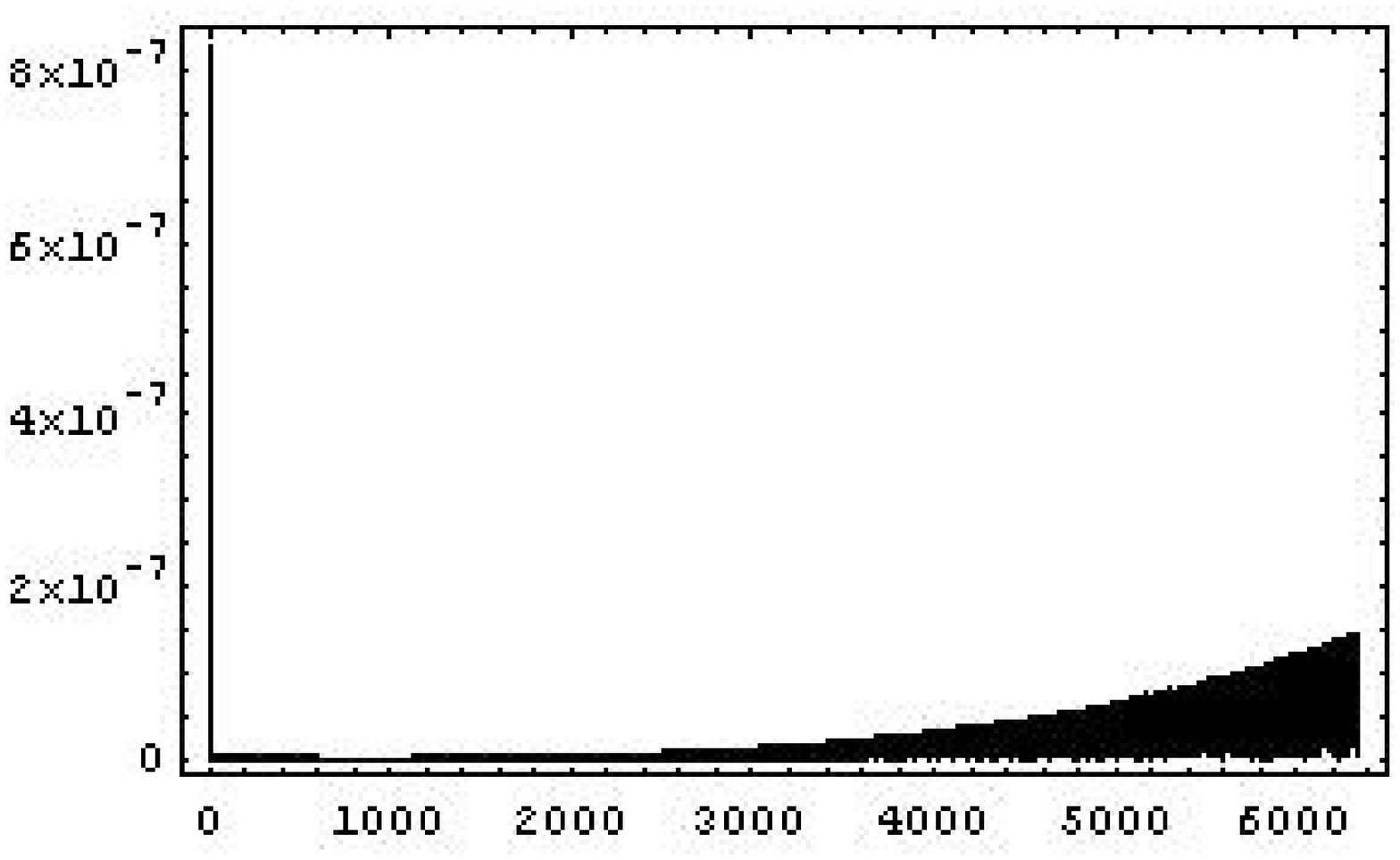}
\caption{{\footnotesize Graphic $(d\Omega_{GW}/dln\nu)$ X
$(\nu\times 10^{-18}Hz)$ for the model where  $\bar{A}=0.5$,
$\alpha=1$, $\Omega_{c_0}=0.96$ and $\Omega_{m_0}=0.04$}}
\label{espectro_10}
\end{minipage} \hfill
\begin{minipage}[b]{0.47\linewidth}
\includegraphics[width=\linewidth]{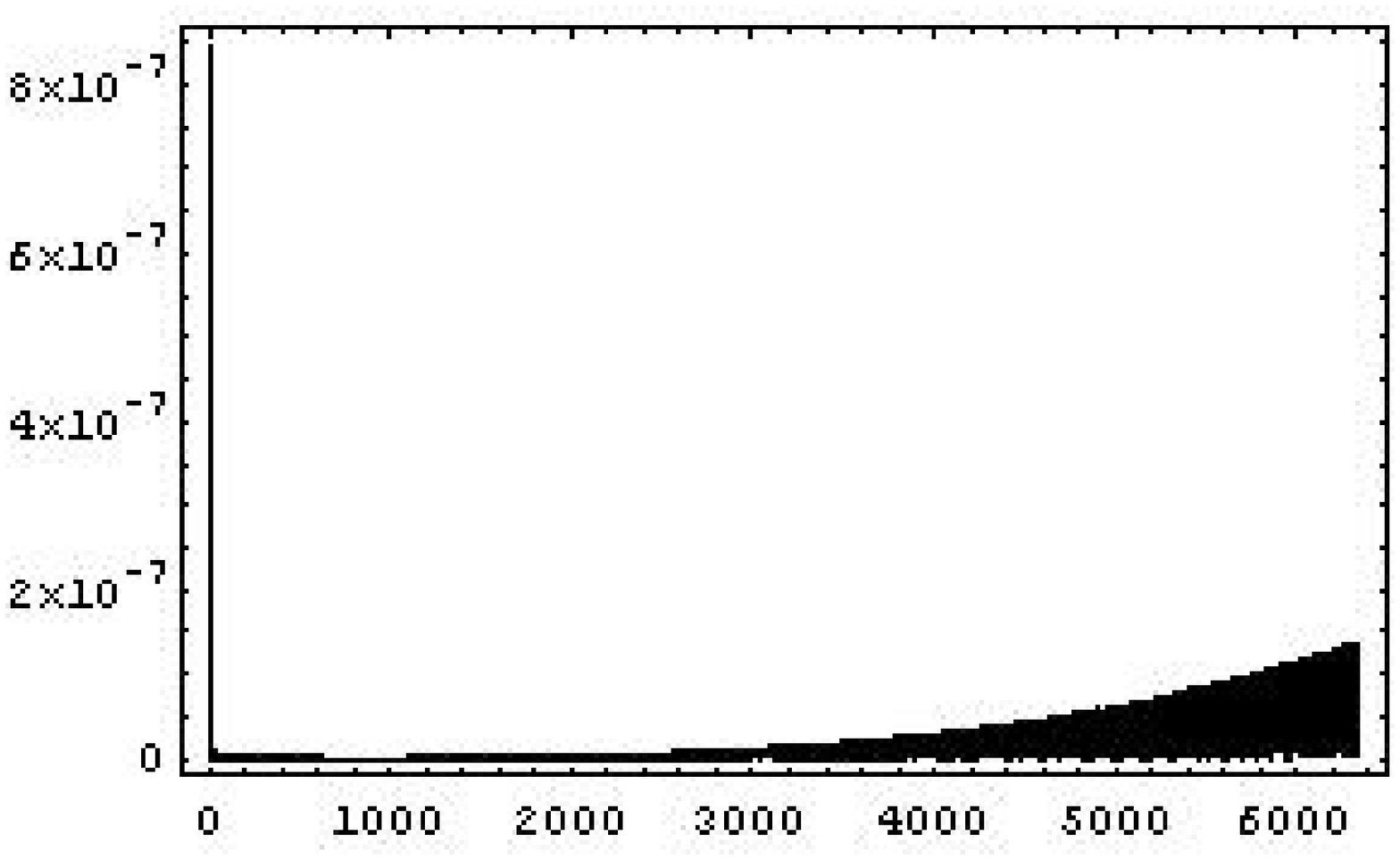}
\caption{{\footnotesize Graphic $(d\Omega_{GW}/dln\nu)$ X $(\nu\times
10^{-18}Hz)$
for the model where $\bar{A}=0.5$, $\alpha=1$,
$\Omega_{c_0}=0.7$ and $\Omega_{m_0}=0.3$}} \label{espectro_11}
\end{minipage} \hfill
\end{figure}

\begin{figure}[!h]
\begin{minipage}[b]{0.47\linewidth}
\includegraphics[width=\linewidth]{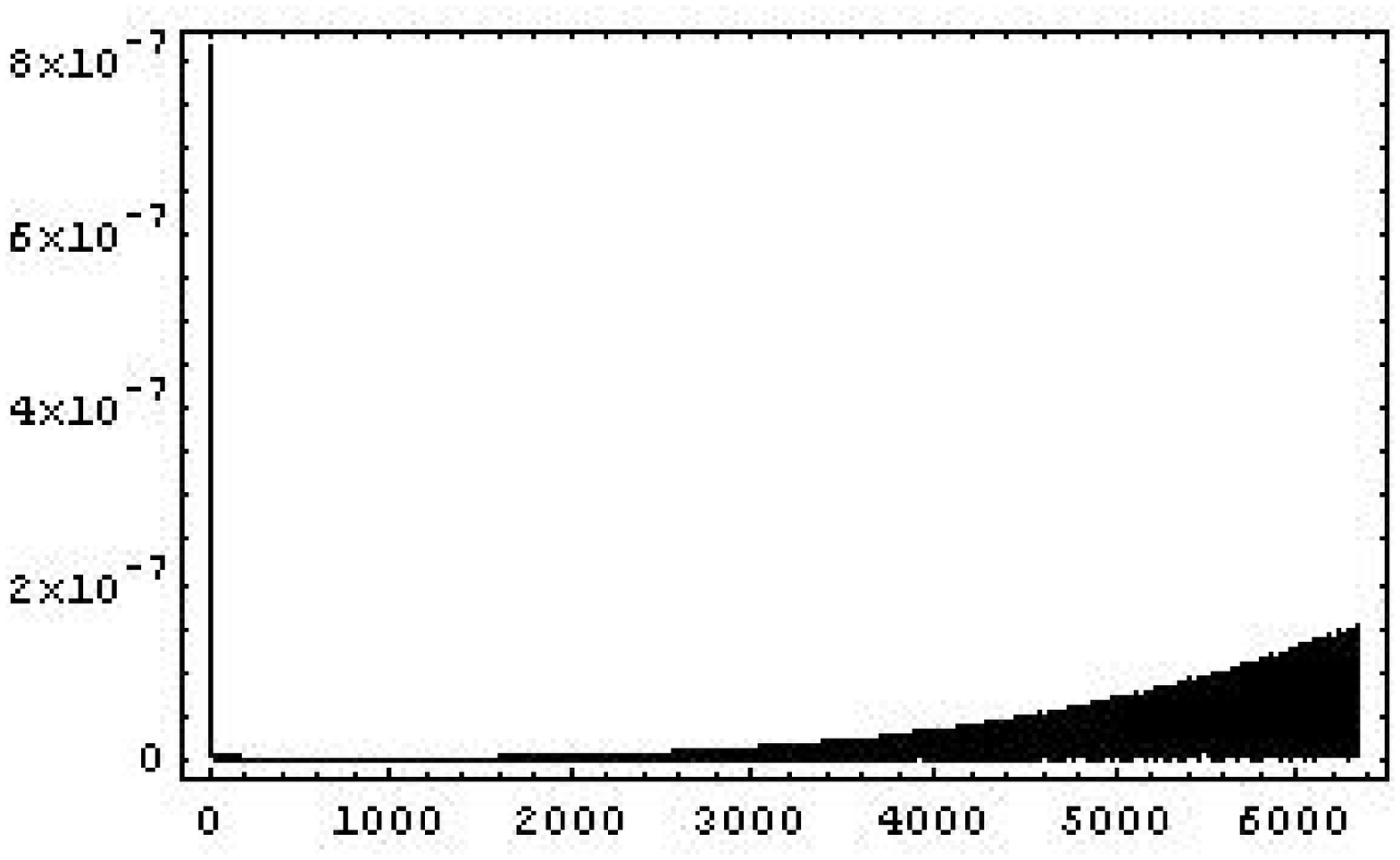}
\caption{{\footnotesize Graphic $(d\Omega_{GW}/dln\nu)$ X $(\nu\times
10^{-18}Hz)$
for the model where $\bar{A}=0.5$, $\alpha=0.5$,
$\Omega_{c_0}=0.96$ and $\Omega_{m_0}=0.04$}} \label{espectro_12}
\end{minipage} \hfill
\begin{minipage}[b]{0.47\linewidth}
\includegraphics[width=\linewidth]{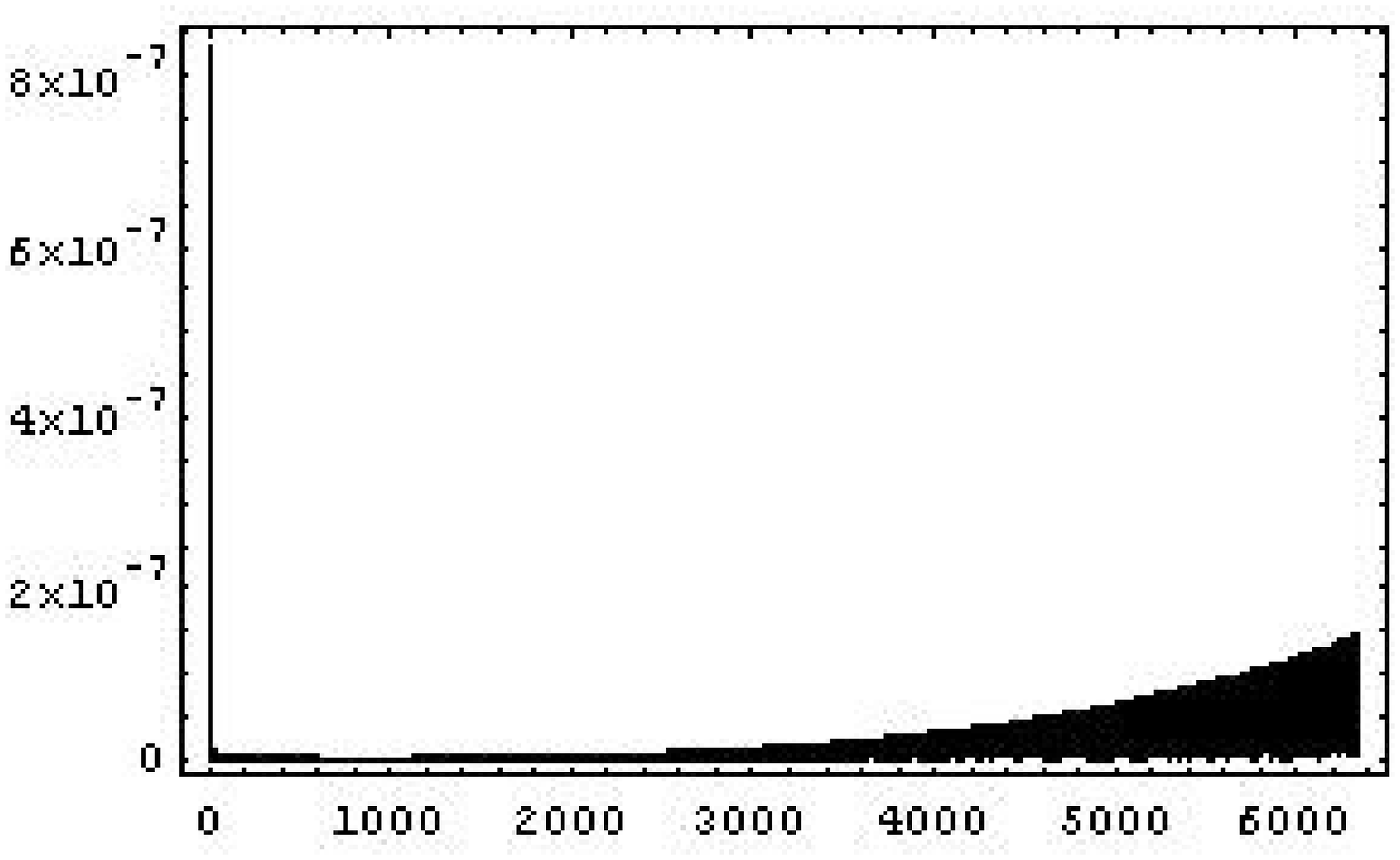}
\caption{{\footnotesize Graphic $(d\Omega_{GW}/dln\nu)$ X $(\nu\times
10^{-18}Hz)$
for the model where $\bar{A}=0.5$, $\alpha=0.5$,
$\Omega_{c_0}=0.7$ and $\Omega_{m_0}=0.3$}} \label{espectro_13}
\end{minipage} \hfill
\end{figure}

\begin{figure}
\begin{minipage}[b]{0.47\linewidth}
\includegraphics[width=\linewidth]{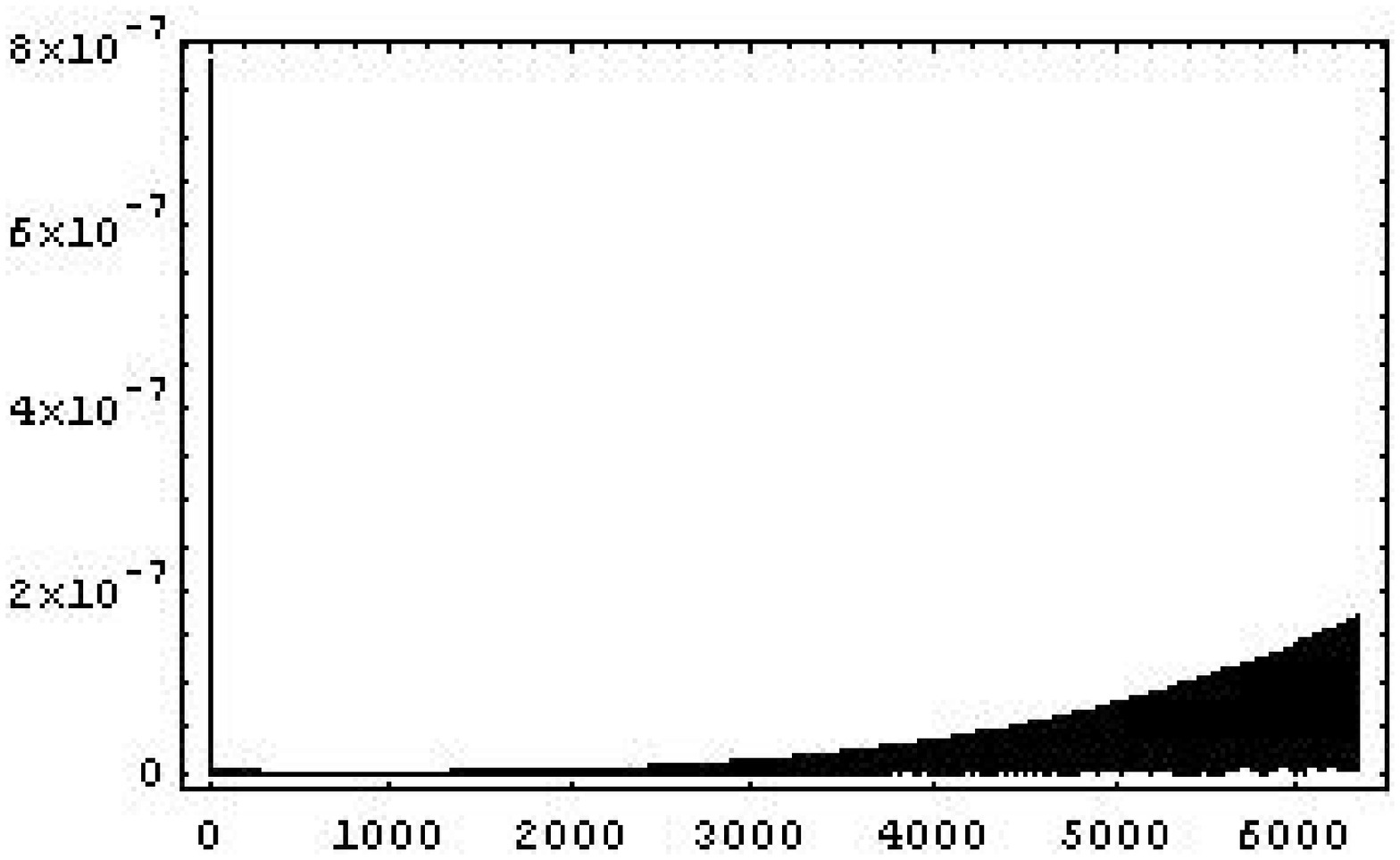}
\caption{{\footnotesize Graphic $(d\Omega_{GW}/dln\nu)$ X $(\nu\times
10^{-18}Hz)$
for the model where $\bar{A}=0.5$, $\alpha=0$,
$\Omega_{c_0}=0.96$ and $\Omega_{m_0}=0.04$}} \label{espectro_17}
\end{minipage} \hfill
\begin{minipage}[b]{0.47\linewidth}
\includegraphics[width=\linewidth]{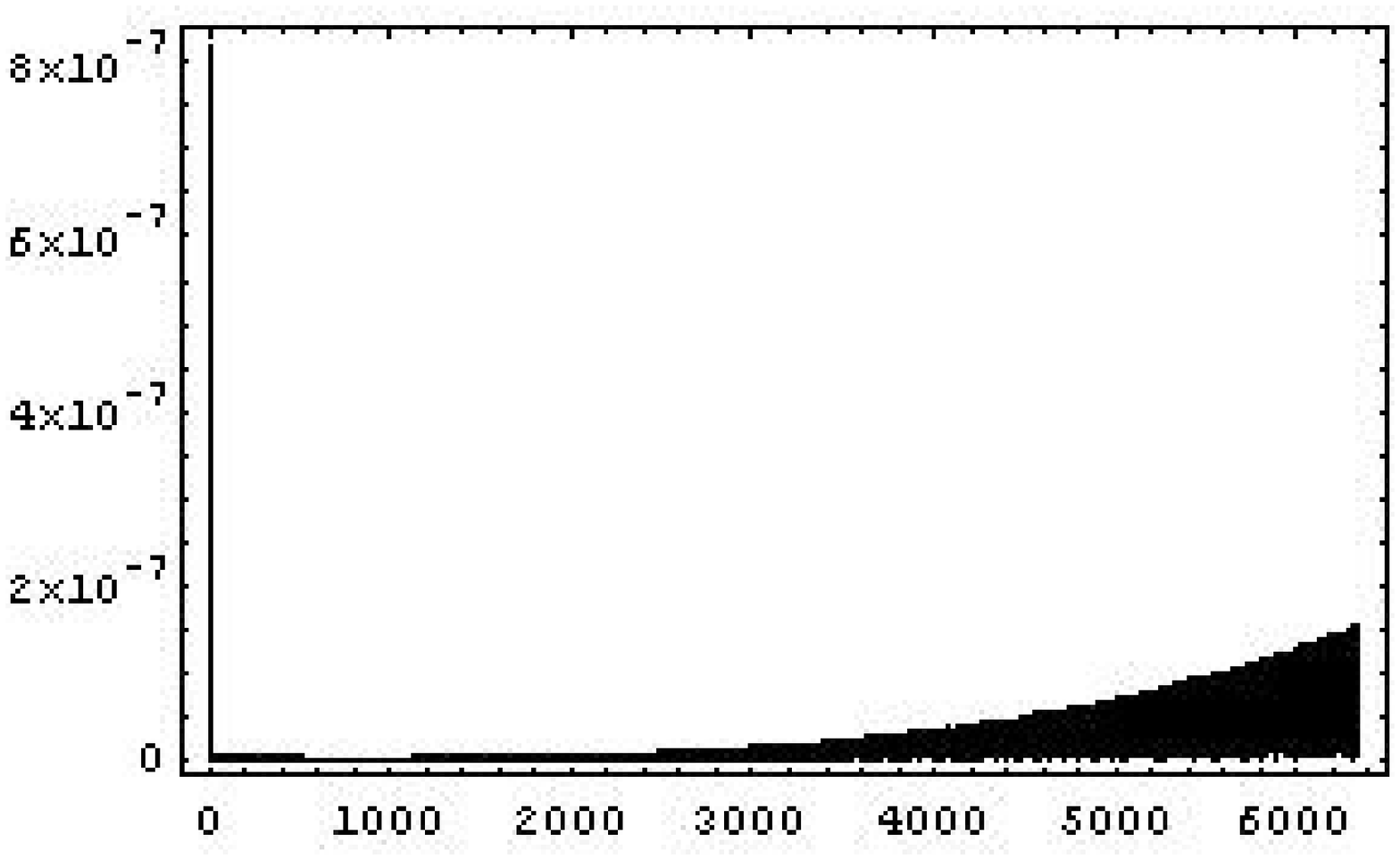}
\caption{{\footnotesize Graphic $(d\Omega_{GW}/dln\nu)$ X $(\nu\times
10^{-18}Hz)$
for the model where $\bar{A}=0.5$, $\alpha=0$,
$\Omega_{c_0}=0.7$ and $\Omega_{m_0}=0.3$}} \label{espectro_18}
\end{minipage} \hfill
\end{figure}

For each of the cases of interest found in table
(\ref{table_models}), we have assigned some common parameters,
namely the initial conditions $h(z_i)=\nu^{-3} 10^{-5}$,
$z_i=4000$; the range of frequencies $10^{-18} Hz \leq \nu \leq
10^{-15} Hz$, and the normalization constant imposed by the
constraint equation (\ref{constr}). The factor $\nu^{-3}$ in the
initial conditions fix the primordial spectrum. The resulting
spectra are presented in figures
(\ref{espectro_5}-\ref{espectro_18}).

Figure (\ref{espectro_5}) refers to the pure pressureless fluid
case and one can see that for $\nu \to 0$ the amplitude reaches
its maximum. It decreases very quickly and, for greater values of
$\nu$, it increases slowly.

On the other hand, the spectrum corresponding to a cosmological
constant dominated Universe (\ref{espectro_14}) has a different
behavior. The amplitude grows faster as the frequency increases
and reaches a value six times greater than the previous one. Figures
(\ref{fig_5}) and (\ref{fig 1}) shows the behaviour of the spectrum
for very low frequencies for the $\Lambda$-CDM case. For the other models,
the spectrum is very similar.
Still referring to the $\Lambda$-CDM model, figure
(\ref{espectro_15}) shows a behavior similar to the one from the
pressureless fluid, but the amplitude growing is faster.

The other graphics correspond to the Chaplygin gas and present a
similar shape. However, in general the Chaplygin gas models lead
to a smaller amplitude in the gravitational wave spectra. This
amplitude is still smaller for the case of pure dark matter model.
Its also important to remark that figures
(\ref{espectro_10}-\ref{espectro_11}) refers to the Chaplygin gas
in its traditional form (with the equation of state $p=-A/\rho$,
since $\alpha=1$) while the others correspond to the generalized
Chapligyn gas.

\section{Conclusions}
\label{conclusions}

In this work we have studied the fate of gravitational waves in the context of
cosmological models where dark energy is described by the generalized Chaplygin gas.
We have exploited the cases where dark energy and dark matter are unified by
the Chaplygin gas, as well as the cases where a pressureless dark component is present besides
the Chaplygin gas. In all situations, however, we have taken into account the baryonic component as infered from the primordial nucleosynthesis.
We perform also an analysis, at same time analytical and numerical, of the
$\Lambda$-CDM model and of a pure dark matter model (where pressureless matter
is the only component of the Universe). In this sense, the goal of the present work was to try to identify special
signature of each model in the energy spectrum as function of the frequency.
\par
The spectra for all models exhibit the same shape: initially the energy decreases with
frequency, becomes almost constant, and then it increases with frequency. However, from the results it comes out that the $\Lambda$-CDM model presents a spectra with an amplitude greater
then the pure dark matter model by a factor of $6$. The Chaplygin gas model interpolates
these two cases. In a typical situation, for example when $\alpha = 1$, $\bar A = 0.5$ and there are only Chaplygin bas and baryons, the amplitude of the spectra is about $4$ times
smaller than in the $\Lambda$-CDM case. As the $\Lambda$-CDM case is approached
($\bar A \rightarrow 1$), the amplitude increases, but in a very low rate, exceptly at
the neighborhood of $\bar A = 1$.
\par
These results show that gravitational waves discriminate very poorly the different models.
Somehow, this is natural since the gravitational waves equation is sensitive essentially
to the background behavior, in contrast with density perturbations which depends strongly
on the kind of matter content.
This is more accentuated in the comparison of different generalized Chaplygin gas models,
that is, models with different $\alpha$. In fact, the Chaplygin gas models are more sensitive
to the parameter $\bar A$ then $\alpha$.
\par
However, it must be stressed that even if the different models
based on the generalized Chaplygin gas may be quite degenerate in
what concerns the behaviour of gravitational waves, it is possible
to distinguish them from the $\Lambda$-CDM model by analyzing the
typical amplitude of the spectra. But, a complete study of the
value of this amplitude asks for a complementation of the analysis
here done by introducing the primordial spectrum of gravitational
waves, for example, that one coming from the primordial
inflationary phase, which allows to fix preciselly the amplitude
and the initial power spectrum \cite{grisha}.
\par
\vspace{0.5cm}
{\bf Acknowledgements:} We thank CNPq (Brazil) for partial financial support.


\begin{thebibliography}{99}

\bibitem{riess} A.G. Riess et al, Astron. J. {\bf 116}, 1009(1998);
\bibitem{permultter} S. Perlmutter et al., Nature, {\bf 391}, 51(1998);
\bibitem{tonry} J.L. Tonry et al., Astrophys. J. {\bf 594}, 1(2003);
\bibitem{wmap} M. Tegmark et al., {\it Cosmological parameters from SDSS and WMAP},
astro-ph/0310723;
\bibitem{cc} V. Sahni, {\it Dark matter and dark energy}, astro-ph/0403324;
\bibitem{quint1}  I. Zlatev, L. Wang and P.J. Steinhardt, Phys. Rev. Lett. {\bf 82},  896(1999);
\bibitem{quint2}  C. Kolda and D.H. Lyth, Phys. Lett. {\bf B458}, 197
(1999);
\bibitem{pasquier} A. Kamenshchik, U. Moschella and V. Pasquier, Phys. Lett. {\bf B511}, 265(2001);
\bibitem{berto}  M.C. Bento, O. Bertolami and A.A. Sen, Phys. Rev. \textbf{%
D66}, 043507 (2002);
\bibitem{fabris} J.C. Fabris, S.V.B. Gon\c{c}alves and P.E. de Souza, Gen. Rel. Grav.
{\bf 34}, 53(2002);
\bibitem{hoppe} M. Bordemann and J. Hoppe, Phys. Lett. {\bf B317}, 315(1993);
\bibitem{ogawa} N. Ogawa, Phys. Rev. {\bf D62}, 085023(2000);
\bibitem{jackiw} R. Jackiw, {\it A particle field theorist's
lectures on supersymmetric, non-abelian fluid mechanics and d-branes}, physics/0010042;
\bibitem{paul1} R. Crittenden, R.L. Davis and P.J. Steinhardt, Astrophys. J. {\bf 417}, L13(1993);
\bibitem{paul2} R. Crittenden, J.R. Bond, R.L. Davis, G. Efstathiou and P.J. Steinhardt,
Phys. Rev. Lett. {\bf 71}, 324(1993);
\bibitem{kogut} A. Kogut et al., Astrophys. J. Suppl. {\bf 148}, 161(2003);
\bibitem{kovac} J.M. Kovac et al., Nature {\bf 420}, 772(2002);
\bibitem{uzan} A. Riazuelo and J.P. Uzan, Phys. Rev. {\bf D62}, 083506(2000);
\bibitem{ungarelli} A. Buonanno, M. Maggiore and C. Ungarelli, Phys. Rev. {bf D55}, 3330(1997);
\bibitem{sanchez} M.P. Infante and N. S\'anchez, Phys. Rev. {\bf D61}, 083515(2000);
\bibitem{gasperini} M. Gasperini, Phys. Rev. {\bf D56}, 4815(1997);
\bibitem{Lineweaver} C.H. Lineweaver, \textit{Cosmological parameters},
astro-ph/0112381, talk presented at COSMO-01, Rovaniemi, Finland,
August 29-September 4(2001);
\bibitem{weinberg} S. Weinberg, {\bf Gravitation and cosmology}, Wiley, New York (1972);
\bibitem{lifschitz} E. M. Lifschitz and I. Khalatnikov, Adv. Phys. {\bf 12}, 185(1963);
\bibitem{grisha} L.P. Grishchuk, Lect. Notes Phys. {\bf 562}, 167(2001).


\end{thebibliography}
\end{document}